# Collusion resistant self-healing key distribution in mobile wireless networks

## Ratna Dutta*

Department of Mathematics,
Indian Institute of Technology Kharagpur,
Kharagpur 721302, India
Email: ratna@maths.iitkgp.ernet.in
*Corresponding author

## Sugata Sanyal

School of Technology & Computer Science,
Tata Institute of Fundamental Research,
Mumbai 400005, India
Email: sanyal@tifr.res.in

**Abstract:** A fundamental concern of any secure group communication system is key management and wireless environments create new challenges. One core requirement in these emerging networks is self-healing. In systems where users can be offline and miss updates, self-healing allows a user to recover lost session keys and get back into the secure communication without putting extra burden on the group manager. Clearly, self-healing must only be available to authorised users. This paper fixes the problem of collusion attack in an existing self-healing key distribution scheme and provides a highly efficient scheme as compared to the existing works. It is computationally secure, resists collusion attacks made between newly joined users and revoked users and achieves forward and backward secrecy. Our security analysis is in an appropriate security model. Unlike the existing constructions, our scheme does not forbid revoked users from rejoining in later sessions.

**Biographical notes:** Ratna Dutta is currently an Assistant Professor in the Department of Mathematics, Indian Institute of Technology, Kharagpur, India. She received her PhD in Computer Science from Indian Statistical Institute, India, in 2006. She visited ENSTA, Paris, in 2006 and worked as an associate scientist at the Institute for Infocomm Research, Singapore during 2006–2008. She was also a post-doc at the Claude Shannon Institute, Department of Computer Science, National University of Ireland, Maynooth, during 2008–2010. Her research interests include public key cryptography, elliptic curve cryptosystem and pairings, dynamic group key agreement, password-based protocols, and wireless ad hoc networks.

Sugata Sanyal is a Professor in the School of Technology & Computer Science at the Tata Institute of Fundamental Research. He has worked in diverse areas of computer architecture, parallel processing, fault tolerance and coding theory and in the area of security. He is in the editorial board of many international journals, and is collaborating with scientists from India and abroad.

## 1 Introduction

Mobile wireless ad hoc networks have wide applications in military operations, rescue missions and scientific explorations, where there are usually no network infrastructure support and the adversary may intercept, modify, and/or partially interrupt the communication. In such applications, security becomes a critical concern. Key distribution and subsequent key updating are cornerstones for secure communication in such networks. The life of such communication network is usually partitioned into short time-periods called sessions. A group manager generates and distributes a session key to the users in the communication group at initialisation stage. All data broadcast within the group should be encrypted with the session key so that only authorised users with session key can access the messages. The session key must be updated with sessions upon each membership change for secure communication.

The traditional approaches for key distribution and group re-keying used for reliable network are not suitable for large and dynamic wireless networks. Key distribution over reliable channels requires strong infrastructure such as wired network and a lot of bandwidth for delivering data punctually to destination. Key distribution over unreliable channels confronts more constraints and challenges. Key distribution messages can be delayed when they are delivered and some messages might never reach to some authorised users. Request or re-transmission between individual users and the group manager in a large group will induce much communication overhead and make the group manager heavily burdened. These are infeasible in many multi-media distribution systems that are based on a uni-directional broadcast distribution channel (Safavi-Naini and Wang, 2000; Naor et al., 2001; Perrig et al., 2001; Gada et al., 2004; Vasudevan and Sanyal, 2004).

Self-healing key distribution is a potential candidate to establish session keys for secure communication to large and dynamic groups in highly mobile, volatile and potentially hostile wireless networks, where frequent membership changes may be necessary and the ability to revoke users during certain exchanges is desirable. In such situations the session keys need to be used for a short time-period or need to be updated frequently. Self-healing is a good property for key distribution in wireless mobile and ad hoc networks, where the nodes/devices are powered by batteries and have the unique feature of moving in and out of range frequently. There might be situations where some users are not constantly online or experience burst packet losses. It can rejoin the group once the power is on again. All these factors can take great advantage from self-healing key distribution schemes with revocation capability.

## 1.1 Self-healing key distribution

The main concept of self-healing key distribution schemes is that users, in a large and dynamic group communication over an unreliable network, can recover lost session keys on their own, even if they have lost some previous key distribution messages, without requesting additional transmissions from the group manager. This reduces network traffic and the risk of user exposure through traffic analysis and alleviates the burden on the group manager. The key idea of self-healing key distribution schemes is to broadcast information that is useful information only to trusted members. Combined with users' pre-distributed secrets, this broadcast information enables a trusted member to reconstruct a shared key. However, a revoked member is unable to infer useful information from the broadcast. The only requirement that a user must satisfy to recover the lost keys through self-healing is its membership in the group both before and after the sessions in which the broadcast packet containing the key is sent. A user who has been offline for some period is able to recover the lost session keys immediately after coming back online. Thus the self-healing approach to key distribution is stateless.

## 1.2 Related work

Broadcast encryption is a closely related area which has received much attention from both the network and cryptography community. Efficient key distribution and key management mechanisms are at the core of this. The area of broadcast encryption was formally defined by Fiat and Naor (1994) after the work of Berkovit (1991) and has been extensively studied since then. A number of approaches have been proposed: re-keying schemes for dynamic groups, broadcast schemes with tracing capability, users revocation from a predefined subset of users etc. A few of them are by Abraham et al. (2004), Blundo et al. (1993), Dal et al. (2008), Dey et al. (2007), Just et al. (1994), Tiwari et al. (2007). However, the underlying networks are assumed to be reliable in all the above works. Self-healing key distribution schemes are introduced by Staddon et al. (2002). They provide formal definitions and security notions that are later generalised by Liu et al. (2003) and Blundo et al. (2004). Staddon et al. (2002) discuss lower bounds on the resources required on the schemes and design some constructions. The constructions given by Staddon et al. (2002) suffer from high storage and communication overhead. Liu et al. (2003) introduce a novel personal key distribution scheme and by combining it with the self-healing technique in the work of Staddon et al. (2002), they propose new constructions that improve the storage and communication overhead greatly. They provide two techniques that allow a trade-off between the broadcast size and the recoverability of lost session keys. These two methods further reduce the broadcast message size – one in the case where there are frequent but short-term disruptions of communication and the other in the case where there are long-term but infrequent disruptions of communication. Following these pioneering works, a number of self-healing key distribution approaches are proposed (Blundo et al., 1996; Staddon et al., 2002; Liu et al., 2003; More et al., 2003; Saez, 2004; Hong and Kang, 2005; Saez, 2005) to achieve unconditional security in formal generalised model with improved efficiency. Blundo et al. (2004) show an attack to the first construction by Staddon et al. (2002) and develop a new self-healing technique different from Staddon et al. (2002) under a slightly modified framework. They present a new mechanism for implementing the self-healing approach, extend the self-healing approach to key distribution and propose a scheme which enables a user to recover all keys associated with the sessions in which it is member of the communication group from a single broadcast message. Blundo et al. (2004) analyse the definitions proposed by Staddon et al. (2002) and Liu et al. (2003) and show that some of the security requirements stated therein, cannot be achieved by any protocol. They propose a new definition of self-healing key distribution and propose constructions that are proven to be secure in that new security framework. Subsequently, they provide some lower bounds on the resources required for implementing such schemes and prove that some of the bounds are tight. Hong and Kang (2005) propose self-healing key distribution constructions having less storage and communication complexity. More et al. (2003) use a sliding window to correct the inconsistent robustness in Staddon et al. (2002).

Further improvements in efficiency are obtained by relaxing the security slightly – from unconditional to computational (Zhang et al., 2003; Zou and Dai, 2006; Dutta et al., 2007; Jiang et al., 2007; Kausar et al., 2007; Dutta et al., 2009). The schemes (Saez, 2004; Saez, 2005; Tian et al., 2008) are based on vector space access structure instead of Shamir's (1979) secret sharing. The hash chain based schemes (Dutta et al., 2007; Jiang et al., 2007; Kausar et al., 2007) are computationally secure and are highly efficient compared to the existing unconditionally secure schemes. However, these hash chain based constructions have the fatal defect of not being collusion resistant in the sense that the collusion between newly joined users and the revoked users are able to recover all the session keys which they are not entitled to. Among the collusion resistance self-healing key distribution schemes (e.g. Blundo et al., 2004; Saez, 2004; Saez, 2005; Tian and He, 2008; Tian et al., 2008) only by Tian et al. (2008) is hash chain based and uses the same self-healing mechanism as introduced by Dutta et al. (2007).

### 1.3 Our contribution

Security, efficiency and scalability are three major evaluation measurements for self-healing key distribution schemes. Besides forward secrecy and backward secrecy, collusion resistance property needs to be addressed. Collusion attacks are dangerous to wireless network key distribution schemes as some members of wireless networks may be revoked regularly. From efficiency point of view, reducing communication overhead is a main concern as energy consumed for computation much depends on algorithm and hardware, and is order of magnitude less than that required for communication (Carman et al., 2000). Also it is required to give consideration to the resource-limited property of nodes while designing protocols for wireless networks. Scalability is another issue as the wireless network may scale up to thousands of nodes and operations are required to finish in a timely manner despite of frequent change of node topology and density.

We address the problem of introducing collusion resistance property to the self-healing key distribution scheme proposed by Dutta et al. (2008). Collusion attacks would cause serious damages to key distribution schemes as some users of wireless networks may be revoked regularly. We use vector space secret sharing together with one-way hash function. Vector space secret sharing helps to realise general monotone decreasing access structure for the family of subsets of users that can be revoked instead of a threshold one. One way hash chain contributes to reduce communication overheads. In our design, a user is assigned a pre-arranged life cycle by the group manager during user's set-up phase and is revoked once its life cycle finishes.

We achieve the following unique features in our designs as compared to the existing similar schemes:

a  Our scheme is anti-collusive in the sense that it can resist collusion between the newly joined users and the revoked users, together with forward and backward secrecy.

b  Our scheme realises a flexible access structure and achieves scalability.

c  Our scheme allows revoked users to rejoin in later sessions with new identities, while this rejoining is prohibited for most of the existing hash chain based self-healing key distribution schemes.

d  Our scheme is communicationally more efficient than the existing schemes as history of revoked users is not sent as part of the broadcast message. Also storage overhead is less as compared to the previous constructions.

A user may get compromised and need a rapid revocation from the group by the group manager. Thus the group manager has to keep track of compromised users using some traitor tracing algorithm, which might be expensive. In our setting, the group manager pre-selects the session of revocation for a user during user's set-up phase by assigning the user a pre-arranged life cycle. The user is revoked from the system by the group manager once its life cycle is over irrespective of user gets compromised or not. The joining session can be selected by the user. Our designs allow a revoked user to join at a later session with new identity and a new life cycle starting from its new joining session. The group manager believes that a user behaves honestly and will not get compromised during its life cycle. Therefore, there is no need for the group manager to use expensive traitor tracing algorithms in handling compromised nodes. The selection of a user's life cycle is completely determined by the group manager.

Assigning each user a pre-arranged life cycle by the group manager and not allowing the user to revoke before its life cycle completes, has natural appeal in many applications. Several innovative business models allow contractual subscription or rental by the service provider for the scalability of business and do not allow the user to get revoked before his contract is terminated. Our scheme is suitable for such applications. Moreover, rejoining of revoked users can be done in our scheme at later sessions with new identities without compromising security, unlike the existing self-healing schemes.

## 2 Preliminaries

We begin by explaining key distribution problem and self-healing property, following it provide definition of one-way function and generalised secret sharing schemes using access structures, and finally we briefly define the security model for self-healing key distribution. The following notations are used throughout the paper (Table 1).

### 2.1 Key distribution and self-healing

Consider the scenario which has a set-up for pay-per-view TV channel. Suppose $\{U_1,…,U_n\}$ is a dynamically changing group of users (clients) and $GM \notin \{U_1,…,U_n\}$ is the group manager (the cable operator). The problem is how the GM can securely communicate with its dynamically changing group of clients over an insecure broadcast channel, so that only authorised clients (who pay) may view the content broadcast by the GM. The GM encrypts the content using a session key. We need a mechanism of distributing this session key in such a way that only the authorised users can recover this session key and decrypt the encrypted content. This mechanism is referred to as the key distribution problem. Our goal is to minimise the overhead for this key distribution keeping the following issues in mind: (a) group re-keying is needed on each membership change; (b) depending on specific nature of applications, we can adopt periodic group re-keying; (c) efficient and secure revocation as well as joining mechanisms are required for dynamic groups etc.

On top of this, $U_i$ may get offline for some time due to power failure and may need to recover lost session keys immediately after being online. Self-healing property enables qualified users to recover lost session keys on their own, without requesting additional transmission from the GM.

**Table 1** Notations used in the paper

| | |
|---|---|
| $\mathcal{U}$ | Set of all users in the networks |
| $U_i$ | $i$-th user |
| GM | Group Manager |
| $n$ | Total number of users in the network |
| $m$ | Total number of sessions |
| $t$ | The maximum number of compromised user |
| $F_q$ | A field of order q |
| $S_i$ | Personal secret of user $U_i$ |
| $SK_j$ | Session key generated by the GM in session j |
| $\mathcal{B}_j$ | Broadcast message by the GM during session j |
| $Z_{i,j}$ | The information learned by $U_i$ through $\mathcal{B}_j$ and $S_i$ |
| $R_j$ | The set of all revoked users in session $j$ |
| $\mathcal{H}$ | A cryptographically secure one-way function |
| $S^B$ | Backward key seed generated by the GM |
| $K_i^B$ | $i$-th backward key in the backward key chain |

## 2.2 One-way function

Our constructions for self-healing key distribution are based on the intractability of one-way function. Informally speaking, a one-way function $f : A \to B$ satisfies the following two properties where $A$ and $B$ are two finite set: (a) $f$ is easy to compute; and (b) $f^{-1}$ is hard to invert, i.e. it is difficult to get $x$ from $f(x)$. See Goldreich (2001) for a formal definition of one-way function. An important component of our system is a cryptographically secure one way Hash function. The underlying principle here is that we must have a measure of the difficulty of reversing such functions. More formally a function $H : A \to B$ is a cryptographically secure hash function if it satisfies the following requirements:

- $\mathcal{H}$ can be applied to any size input and produce a fixed length output.
- $\mathcal{H}$ is easy to compute.
- $\mathcal{H}$ has the one-way property, i.e. Given $\mathcal{H}(x)$ it is computationally infeasible to find $x$.
- $\mathcal{H}$ is weak collision resistant, i.e. Given $x$ it is computationally infeasible to find $y \neq x$ with $\mathcal{H}(y) = \mathcal{H}(x)$.
- $\mathcal{H}$ is strong collision resistant, i.e. it is computationally infeasible to find a distinct pair $(x, y)$ with $\mathcal{H}(x) = \mathcal{H}(y)$.

In what follows $A$ and $B$ are $GF(q)$. As the hash function landscape is constantly changing we do not specify a particular algorithm to compute $\mathcal{H}$, but note that our construction is not dependent on a particular hash function.

## 2.3 Cryptographically Secure Pseudo Random Bit Generators (CSPRBG)

Our system requires a good supply of 'random' numbers. In most practical environments the generation of random numbers is inefficient and the storage and distribution of the resulting random numbers is impractical. In such situations random number generators are replaced by Pseudo Random Number Generators (PRNG). Also without loss of generality we can consider Pseudo Random Bit Generators (PRBG). A PRBG is a deterministic algorithm that inputs a random binary sequence called a seed and outputs a longer binary stream that appears random. The resulting sequence is not random but can be tested in order to gauge predictability.

One such test is the *next-bit test*. A PRBG is said to pass the next-bit test if there is no polynomial time algorithm which, on input of the first $l$ bits of an output sequence $s$, can predict the $(l+1)$)th bit of $s$ with probability greater than 0.5. A PRBG that passes this test, possibly under a plausible security assumption such as the discrete log problem, is referred to as a Cryptographically Secure Pseudo random bit Generator (CSPRBG).

## 2.4 Secret sharing schemes

In this section we define secret sharing schemes which play an important role in distributed cryptography.

Definition 1 (Access Structure): *Let $\mathcal{U} = \{U_1,...,U_n\}$ be a set of participants. A collection $\Gamma \subseteq 2^{\mathcal{U}}$ is monotone if $B \in \Gamma$ and $B \subseteq C \subseteq \mathcal{U}$ imply $C \in \Gamma$. An access structure is a monotone collection $\Gamma$ of non-empty subsets of $\mathcal{U}$, i.e. $\Gamma \subseteq 2^{\mathcal{U}} \setminus \{\emptyset\}$. The sets in $\Gamma$ are called the authorised sets. A set B is called minimal set of $\Gamma$ if $B \in \Gamma$, and for every $C \subset B$, $C \neq B$, it holds that $C \notin \Gamma$. The set of minimal authorised subsets of $\Gamma$ is denoted by $\Gamma_0$ and is called the basis of $\Gamma$. Since $\Gamma$ consists of all subsets of $\mathcal{U}$ that are supersets of a subset in the basis $\Gamma_0$, $\Gamma$ is determined uniquely as a function of $\Gamma_0$. More formally, we have $\Gamma = \{C \subseteq \mathcal{U} : B \subseteq C, B \in \Gamma_0\}$. We say that $\Gamma$ is the closure of $\Gamma_0$ and write $\Gamma = cl(\Gamma_0)$. The family of non-authorised subsets $\overline{\Gamma} = 2^{\mathcal{U}} \setminus \Gamma$ is monotone decreasing, that is, if $C \in \overline{\Gamma}$ and $B \subseteq C \subseteq \mathcal{U}$, then $B \in \overline{\Gamma}$. The family of non-authorised subsets $\overline{\Gamma}$ is determined by the collection of maximal non-authorised subsets $\overline{\Gamma}_0$.*

Example: *In case of a $(t, n)$-threshold access structure, the basis consists of all subsets of (exactly) t participants, i.e. $\Gamma = \{B \subseteq \mathcal{U} : |B| \geq t\}$ and $\Gamma_0 = \{B \subseteq \mathcal{U} : |B| = t\}$.*

Definition 2 (Secret Sharing): *Let $\mathcal{K}$ be a finite set of secrets, where $\mathcal{K} \geq 2$. An n-party secret sharing scheme $\Pi$ with secret domain $\mathcal{K}$ is a randomised mapping from $\mathcal{K}$ to a* set of n-tuples $S_1 \times S_2 \times \ldots \times S_n$, where $S_i$ is called the share domain of $U_i \in \mathcal{U}$. A dealer $D \notin \mathcal{U}$ distributes a secret $K \in \mathcal{K}$ according to $\Pi$ by first sampling a vector of shares $(s_1, \ldots, s_n)$ from $\Pi(K)$, and then privately communicating each share $s_i$ to the party $U_i$. We say that $\Pi$ realises an access structure $\Gamma \subseteq 2^{\mathcal{U}}$ if the following two requirements hold:

Correctness: *The secret $K$ can be reconstructed by any authorised subset of parties. That is, for any subset $B \in \Gamma$ (where $B = \{U_{i_1}, \ldots, U_{i_{|B|}}\}$), there exists a reconstruction function $Rec_B : S_{i_1} \times \ldots \times S_{i_{|B|}} \to \mathcal{K}$ such that for every $K \in \mathcal{K}$, $Prob[Rec_B(\Pi(K)_B) = K] = 1$, where $\Pi(K)_B$ denotes the restriction of $\Pi(K)$ to its B-entries.*

Privacy: *Every unauthorised subset cannot learn anything about the secret (in the information theoretic sense) from their shares. Formally, for any subset $C \notin \Gamma$, for every two secrets $K_1, K_2 \in \mathcal{K}$, and for every possible shares $\langle s_i \rangle_{U_i \in C}$, $Prob[\Pi(K_1)_C = \langle s_i \rangle_{U_i \in C}] = Prob[\Pi(K_2)_C = \langle s_i \rangle_{U_i \in C}].$*

The above correctness and privacy requirements capture the strict notion of perfect secret sharing, which is the one most commonly referred in the secret sharing literature. Next we define the class of linear secret sharing schemes. There are several equivalent definitions for these schemes (Beimel, 1996), we provide the following.

Definition 3 (Linear Secret Sharing): *Let F be a field. A secret sharing scheme $\Pi$ is said to be linear over F if: (1) The secret domain $\mathcal{K}$ is a subset of F. (2) There exists $d_1, \ldots, d_n$ such that each share domain $S_i$ is a subset of the vector space $F^{d_i}$. (3) The randomised mapping $\Pi$ can be computed as follows. First, the dealer $D \notin \mathcal{U}$ chooses independent random variables, denoted by $r_1, \ldots, r_l$, each uniformly distributed over F. Then, each coordinate of each of the n shares is obtained by taking a linear combination of $r_1, \ldots, r_l$ and the secret $K \in \mathcal{K}$.*

We are interested in a special case of linear secret sharing scheme, namely *vector space secret sharing scheme* that will be described shortly.

Definition 4 (Vector Space Access Structure): *Suppose $\Gamma$ is an access structure, and let $(Z_q)^l$ denote the vector space of all l-tuples over $Z_q$, where q is prime and $l \geq 2$. Suppose there exists a function $\Phi : \mathcal{U} \cup \{D\} \to (Z_q)^l$ which satisfies the property: $B \in \Gamma$ if and only if the vector $\Phi(D)$ can be expressed as a linear combination of the vectors in the set $\{\Phi(U_i) : U_i \in B\}$. An access structure $\Gamma$ is said to be a vector space access structure if it can be defined in the above way.*

We now present vector space secret sharing scheme that was introduced by Brickell (1983).

- *Initialisation*: For $1 \leq i \leq n$, $D$ gives the vector $\Phi(U_i) \in (Z_q)^l$ to $U_i$. These vectors are public.

- *Share Distribution*:
  1. Suppose $D$ wants to share a key $K \in Z_q$. $D$ secretly chooses (independently at random) $l-1$ elements $a_2, \ldots, a_l$ from $Z_q$.
  2. For $1 \leq i \leq n$, $D$ computes $s_i = v.\Phi(U_i)$, where $v = (K, a_2, \ldots, a_l) \in (Z_q)^l$.
  3. For $1 \leq i \leq n$, $D$ gives the share $s_i$ to $U_i$.

- *Key Recovery*: Let $B$ be an authorised subset, $B \in \Gamma$. Then
$$\Phi(D) = \sum_{\{i:U_i \in B\}} \Lambda_i \Phi(U_i)$$
for some $\Lambda_i \in Z_q$. In order to recover the secret $K$, the participants of $B$ pool their shares and computes $\sum_{\{i:U_i \in B\}} \Lambda_i s_i = \sum_{\{i:U_i \in B\}} \Lambda_i v.\Phi(U_i) = v.\left(\sum_{\{i:U_i \in B\}} \Lambda_i \Phi(U_i)\right) = v.\Phi(D) = K \bmod q$.

Thus when an authorised subset of participants $B \in \Gamma$ pool their shares, they can determine the value $K$. On the other hand, one can show that if an unauthorised subset $B \notin \Gamma$ pool their shares, they can determine nothing about the value of $K$ (see Brickell, 1983 for proof).

Example 1: *Shamir's $(t,n)$-threshold scheme can be seen as a special case of the vector space secret sharing scheme, if $q > n$, by defining $l = t$ and choosing $\Phi(U_i) = (1, x_i, x_i^2, \ldots, x_i^{t-1})$ for $1 \leq i \leq n$, where $x_i$ is the x-coordinate given to $U_i$. The resulting scheme is equivalent to the Shamir's scheme.*

Example 2: *Another example of vector space secret sharing concerns access structures that have as a basis a collection of pairs of participants that forms a complete multipartite graph. A graph $G = (V, E)$ is called a complete multipartite graph if the vertex set V can be partitioned into subsets $V_1, \ldots, V_d$ such that $\{x, y\} \in E$ if and only if $x \in V_i$, $y \in V_j$, where $i \neq j$. The sets $V_i$ are called parts. The complete multipartite graph is denoted by $K_{n_1, \ldots, n_d}$ if $|V_i| = n_i, 1 \leq i \leq d$. When $n_1 = n_2 = \cdots = n_d = 1$, we get the complete multipartite graph $K_{1, \ldots, 1}$ (with d parts) which is in fact a complete graph and is denoted by $K_d$. It can be shown that there exists an ideal scheme realising the access structure $cl(E)$ (closure of E) on participant set V by choosing distinct elements $x_1, \ldots, x_d$ of $Z_q$, where $q \geq d$, defining $l = 2$ and choosing $\Phi(U) = (x_i, 1)$ for every participant $U \in V_i$.*

Example 3: *Consider bipartite access structures which are first presented in Padro and Saez (2000). In such a structure $\Gamma$, there is a partition of the set of participants, $\mathcal{U} = X \cup Y$, such that all participants in the same class play an equivalent role in the structure. Any subset $A \subset \mathcal{U}$ is assigned with the point of non-negative integers*
$$\pi(A) = (x(A), y(A)) \in Z^+ \times Z^+,$$

where $x(A) =| A \cap X |, y(A) =| A \cap Y |$ and the structure to a region:

$$\pi(\Gamma) = \{\pi(A) : A \in \Gamma\} \subset Z^+ \times Z^+.$$

To be precise, let $n'$ be the total number of possible real users. Consider a set $\mathcal{U} = X \cup Y$, where $X = \{1,\ldots,n',n'+1,\ldots,n'+t-j-1\}$ contains the $n'$ possible real users and $t-j-1$ dummy users, and $Y = \{n'+t-j,\ldots,n'+t-1\}$ contains $j$ dummy users. So the set $\mathcal{U}$ contains $n = n'+t-1$ users. Let us consider the following bipartite access structure $\Gamma$ defined in $\mathcal{U} = X \cup Y$. $\Gamma = \{A \subset X \cup Y : |A| \geq j+1$ and $|A \cap Y| \geq 1\} \cup \{A \subset X \cup Y : |A \cap Z| \geq t\}$, which corresponds to the following region. $\pi(\Gamma) = \{(x,y) \in Z^+ \times Z^+ : (x \geq t)$ or $(x+y \geq j+1$ and $y \geq 1)\}$. The maximal non-authorised subsets in this structure are defined by points $(t-1,0),(j-1,1),(j-2,2),\ldots,(1,j-1),(0,j)$. Note that non-authorised subsets of a $(t,n)$-threshold structure are defined by $t-1$ users. This bipartite access structure $\Gamma$ cannot be realised by a vector space secret sharing scheme (except in the threshold case $j = t-1$ (see Padro and Saez, 2000 for the details), but by a linear one in which each participant is associated with two vectors instead of one. Therefore, each operation will have twice the cost for the same operation in the threshold case.

Vector space secret sharing scheme is a particular case of linear secret sharing scheme, but where every participant can be associated with more than one vector. Any access structure $\Gamma$ can be realised by a linear secret sharing scheme (Simmons et al., 1991).

## 2.5 Security model

There are two categories of attacks against the existing self-healing key distribution schemes: outside attack and inside attack. The attack launched by users who never participated in the communication group is referred to as the outside attack. The outside attacks are prevented by means of some hardware techniques (e.g. tamper resistance) so that the attacker cannot get personal key from a captured/compromised node. On the contrary, the inside attack is launched by users who ever or will be authorised members of a communication group. However, additional security measures should be provided for the group manager to block the intruders from compromising the group manager.

As to the inside attack, we consider an attack scenario where adversary can compromise more than one user. There are three different scenarios to define adversarial goals depending on their degree of severity. In the first scenario, revoked users collude to acquire the subsequent session keys after they are revoked from the authorised group. Another severe attack is when new users collude to acquire the past session keys before they join the communication group. The most severe attack is when the coalition of both revoked users and new joined users try to acquire all the session keys that they were unauthorised to. We now state the following definitions that model the inside attack. These definitions are aimed to computational security for session key distribution adopting the security model of Liu et al. (2003) and Staddon et al. (2002).

Let $\mathcal{U} = \{U_1,\ldots,U_n\}$ be the universe of the network. We assume the availability of a broadcast unreliable channel and there is a group manager $GM$ who sets up and performs join and revoke operations to maintain a communication group, which is a dynamic subset of users of $\mathcal{U}$. Let $m$ be the maximum number of sessions, and $\mathcal{R} \subset 2^{\mathcal{U}}$ be a monotone decreasing access structure of subsets of users that can be revoked by the group manager GM. Let $i \in \{1,\ldots,n\}$, $j \in \{1,\ldots,m\}$ and $G_j \in \mathcal{U}$ be the group established by the group manager GM in session $j$. In the following definitions, $S_i$ denotes the personal secret of user $U_i$, $SK_j$ is the session key generated by the GM in session $j$, $\mathcal{B}_j$ is the broadcast message by the GM during session $j$, and $Z_{i,j}$ is the information learned by $U_i$ through $\mathcal{B}_j$ and $S_i$.

Definition 5 (Session Key Distribution with Privacy (Staddon et al., 2002)):

1  $\mathcal{D}$ is a session key distribution with privacy if

   a  for any user $U_i \in G_j$, the session key $SK_j$ is efficiently determined from $B_j$ and $S_i$

   b  for any set $R_j \subseteq \mathcal{U}$, where $R_j \in \mathcal{R}$ and $U_i \notin R_j$, it is computationally infeasible for users in $R_j$ to determine the personal key $S_j$

   c  what users $U_1,\ldots,U_n$ learn from $\mathcal{B}_j$ cannot be determined from broadcasts or personal keys alone, i.e. if we consider separately either the set of $m$ broadcasts $\{\mathcal{B}_1,\ldots,\mathcal{B}_m\}$ or the set of $n$ personal keys $\{S_1,\ldots,S_n\}$, then it is computationally infeasible to compute session key $SK_j$ (or other useful information) from either set.

2  $\mathcal{D}$ has $\mathcal{R}$-revocation capability if given any $R_j \subseteq \mathcal{U}$, where $R_j \in \mathcal{R}$, the group manager GM can generate a broadcast $\mathcal{B}_j$ such that for all $U_i \notin R_j$, $U_i$ can efficiently recover the session key $SK_j$, but the revoked users cannot, i.e. it is computationally infeasible to compute $SK_j$ from $\mathcal{B}_j$ and $\{S_l\}_{U_l \in R_j}$.

3  $\mathcal{D}$ is self-healing if the following is true for any $j$, $1 \leq j_1 < j < j_2 \leq m$:

   a  For any user $U_i$ who is a member in sessions $j_1$ and $j_2$, the key $SK_j$ is efficiently determined by the set $\{Z_{i,j_1}, Z_{i,j_2}\}$. In other words, every user $U_i \in G_{j_1}$, who has not been revoked after session $j_1$ and before session $j_2$, can recover all session keys $SK_j$ for $j = j_1,\ldots,j_2$, from the broadcasts $\mathcal{B}_{j_1}$ and $\mathcal{B}_{j_2}$ where $1 \leq j_1 < j_2 \leq m$.

b   Let $1 \leq j_1 < j < j_2 \leq m$. For any disjoint subsets $L_1, L_2 \subset \mathcal{U}$, where $L_1 \cup L_2 \in \mathcal{R}$, the set $\{Z_{l,j}\}_{U_l \in L_1, 1 \leq j \leq j_1} \cup \{Z_{l,j}\}_{U_l \in L_2, j_2 \leq j \leq m}$ cannot determine the session key $SK_j$, $j_1 < j < j_2$, i.e. $SK_j$ cannot be obtained by the coalition $L_1 \cup L_2$, where the set $L_1$ is a coalition of users removed before session $j_1$ and the set $L_2$ is a coalition of users joined from session $j_2$.

Definition 6 ($\mathcal{R}$-wise forward and backward secrecy (Liu et al., 2003)):

1   *A key distribution scheme $\mathcal{D}$ guarantees $\mathcal{R}$-wise forward secrecy if for any set $R_j \subseteq \mathcal{U}$, where $R_j \in \mathcal{R}$, and all $U_s \in R_j$ are revoked before session $j$, it is computationally infeasible for the members in $R_j$ together to get any information about $SK_j$, even with the knowledge of group keys $SK_1, \ldots, SK_{j-1}$ before session $j$.*

2   *A session key distribution $\mathcal{D}$ guarantees $\mathcal{R}$-wise backward secrecy if for any set $J_j \subseteq \mathcal{R}$, where $J_j \in \mathcal{U}$, and all $U_s \in J_j$ join after session $j$, it is computationally infeasible for the members in $J_j$ together to get any information about $SK_j$, even with the knowledge of group keys $SK_{j+1}, \ldots, SK_m$ after session $j$.*

## 3   Our general construction

We consider a setting in which there is a group manager (GM) and $n$ users $\mathcal{U} = \{U_1, \ldots, U_n\}$. All our operations take place in a finite field, $GF(q)$, where $q$ is a large prime number ($q > n$). In our setting, we allow a revoked user to rejoin the group in a later session. Let $\mathcal{H} : GF(q) \to GF(q)$ be a cryptographically secure one-way function. The life of the system is divided in sessions $j = 1, 2, \ldots, m$. The communication group in session $j$ is denoted by $G_j \subset \mathcal{U}$. We consider a linear secret sharing scheme realising some access structure $\Gamma$ over the set $\mathcal{U}$. For simplicity, suppose there exists a public function $\Phi : \mathcal{U} \cup \{GM\} \to GF(q)^l$ satisfying the property $\Phi(GM) \in \langle \Phi(U_i) : U_i \in B \rangle \Leftrightarrow B \in \Gamma$, where $l$ is a positive integer. In other words, the vector $\Phi(GM)$ can be expressed as a linear combination of the vectors in the set $\{\Phi(U_i) : U_i \in B\}$ if and only if $B$ is an authorised subset. Then $\Phi$ defines $\Gamma$ as a *vector space access structure*.

*Set-up*: Let $G_1 \in \mathcal{U}$. The group manager GM chooses independently and uniformly at random $m$ vectors $v_1, v_2, \ldots, v_m \in GF(q)^l$. The group manager randomly picks an initial backward key seed $S^B \in GF(q)$. It repeatedly applies (in the pre-processing time) the one-way function $\mathcal{H}$ on $S^B$ and computes the one-way backward key chain of length $m$:

$$K_i^B = \mathcal{H}(K_{i-1}^B) = \mathcal{H}^{i-1}(S^B) \text{ for } 1 \leq i \leq m.$$

The GM also selects at random $m$ numbers $\beta_1, \ldots, \beta_m \in GF(q)$ by running a PRNG which is cryptographically secure. The $j$-th session key is computed as $SK_j = \beta_j + K_{m-j+1}^B$.

Unlike the existing self-healing key distribution schemes, our setting allows a revoked user to rejoin the group in a later session with a new identity. However, we make the following restriction on the life cycle of each user as determined by the GM. Each user $U_i$ is first assigned a prearranged life cycle $(s_i, t_i)$, where $1 \leq s_i < t_i \leq m$, by the GM, i.e. $U_i$ is involved in $k_i = t_i - s_i + 1$ sessions and is not allowed to revoke before session $t_i$, however $U_i$ may get offline during its life cycle due to power failure. Self-healing is needed at this point. Each user $U_i$, for $1 \leq i \leq n$, receives its personal secret keys corresponding to the $k_i$ sessions $S_i = \{v_{s_i}.\Phi(U_i), \ldots, v_{t_i}.\Phi(U_i); \beta_{s_i}, \ldots, \beta_{t_i}\} \in GF(q)^{2k_i}$ from the group manager via the secure communication channel between them. Here the operation '.' is the inner product modulo $q$.

*Broadcast*: Let $R_j$ be the set of all revoked users for sessions in and before $j$ such that $R_j \notin \Gamma$ and $G_j$ be the set of all non-revoked users in session $j$. In the $j$-th session the GM first chooses a subset of users $W_j \subset \mathcal{U} \setminus G_j$ with minimal cardinality such that $W_j \cup R_j \in \overline{\Gamma}_0$. The GM then computes $Z_j = K_{m-j+1}^B + v_j.\Phi(GM)$ and broadcasts the message

$$\mathcal{B}_j = \{(U_k, v_j.\Phi(U_k)) : U_k \in W_j \cup R_j\} \cup \{Z_j\}.$$

*Session Key Recovery*: When a non-revoked user $U_i$ receives the $j$-th session key distribution message $\mathcal{B}_j$, it recovers $v_j.\Phi(GM)$ as follows: Since $W_j \cup R_j \in \overline{\Gamma}_0$ is the maximal non-authorised subset with minimum cardinality having the property $W_j \in \mathcal{U} \setminus G_j$, the set $B = W_j \cup R_j \cup \{U_i\} \in \Gamma$. Thus $B$ is an authorised subset, and one can write

$$\Phi(GM) = \sum_{\{k:U_k \in B\}} \Lambda_k \Phi(U_k) \qquad (1)$$

for some $\Lambda_k \in GF(q)$. Hence $U_i$ knows $\Lambda_k$ and $v_j.\Phi(U_k)$ for all $k \in B$ and consequently can compute

$$\sum_{\{k:U_k \in B\}} \Lambda_k (v_j.\Phi(U_k)) = v_j.\left(\sum_{\{k:U_k \in B\}} \Lambda_k \Phi(U_k)\right)$$
$$= v_j.\Phi(GM)$$

Then $U_i$ recovers the key $K_{m-j+1}^B$ as $K_{m-j+1}^B = Z_j - v_j.\Phi(GM)$. Finally, $U_i$ computes the current session key $SK_j = \beta_j + K_{m-j+1}^B$.

A user $U_k$ who either does not know its private information $v_j.\Phi(U_k)$ or who is a revoked user in $R_j$, i.e. $U_k \in W_j \cup R_j$, cannot compute $v_j.\Phi(GM)$ because $U_k$ only knows values broadcast in the message $\mathcal{B}_j$ corresponding to an unauthorised subset of the secret sharing scheme. Consequently, $U_k$ cannot recover the backward key $K^B_{m-j+1}$ and hence the $j$-th session key $SK_j$.

*Add Group Members*: When a new user wants to join the communication group starting from session $j$, the user gets in touch with the GM. The GM in turn picks an unused identity $\theta \in GF(q)$, assigns a life cycle $(s_\theta, t_\theta)$ to the new user with $s_\theta = j$, computes the personal secret keys corresponding to $k_\theta = t_\theta - s_\theta + 1$ sessions $S_\theta = \{v_{s_\theta}.\Phi(U_\theta),\ldots,v_{t_\theta}.\Phi(U_\theta); \beta_{s_\theta},\ldots,\beta_{t_\theta}\}$ and gives $S_\theta$ to this new group member via the secure communication channel between them.

**Figure 1** Generation of one-way forward key chain and distribution of users' personal secret keys during the *Set-up* phase of our general construction ($(s_i, t_i)$ being $U_i$'s prearranged life cycle for $1 \leq s_i < t_i \leq m$)

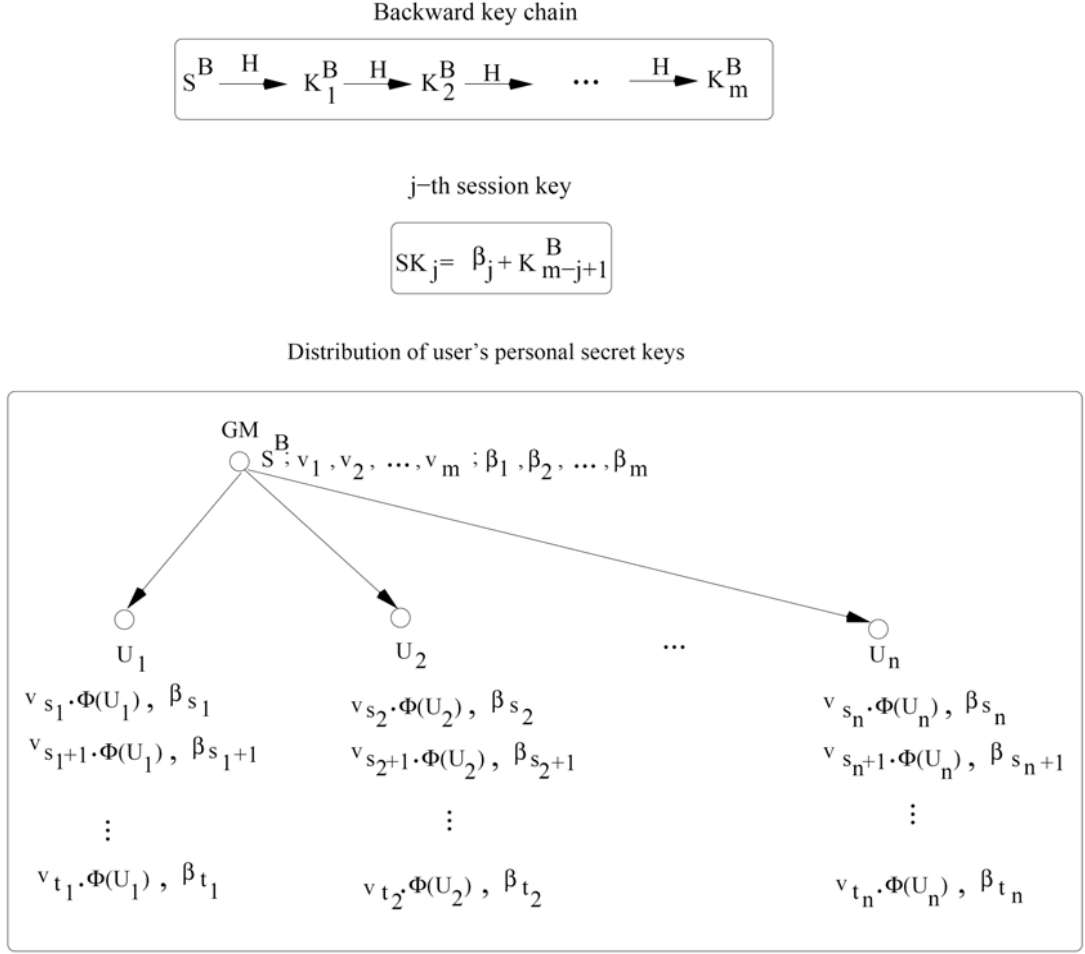

**Figure 2** Broadcast at the $j$-th session, where $U_k \in W_j \cup R_j$

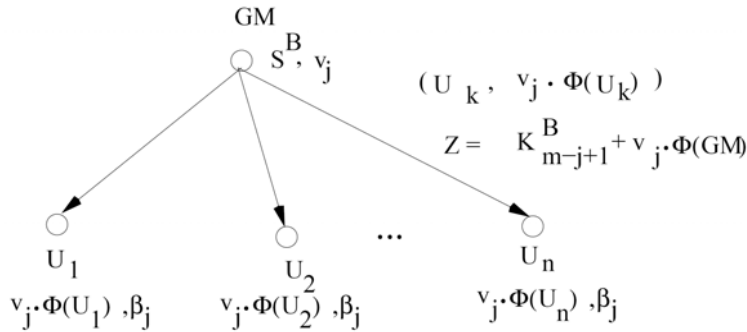

## 4 Self-healing

We now explain our self-healing mechanism in the above constructions: Let $U_i$ be a group member that receives session key distribution messages $\mathcal{B}_{j_1}$ and $\mathcal{B}_{j_2}$ in sessions $j_1$ and $j_2$ respectively, where $1 \leq j_1 \leq j_2$, but not the session key distribution message $\mathcal{B}_j$ for session $j$, where $j_1 < j < j_2$. User $U_i$ can still recover all the lost session keys $SK_j$ for $j_1 < j < j_2$ as desired by point 3(a) of Definition 5 using the following steps:

- $U_i$ recovers from the broadcast message $\mathcal{B}_{j_2}$ in session $j_2$, the backward key $K^B_{m-j_2+1}$ and repeatedly apply the one-way function $\mathcal{H}$ on this and computes the backward keys $K^B_{m-j+1}$ for all $j$, $j_1 \leq j < j_2$.

- $U_i$ then recovers all the session keys $SK_j = \beta_j + K^B_{m-j+1}$, for $j_1 \leq j \leq j_2$.

Note that a user $U_i$ revoked in session $j$ cannot compute the backward keys $K^B_{m-j_1+1}$ for $j_1 > j$. Moreover, since a user is not allowed to revoke before the end of its life cycle, $U_i$ revoked in $j$-th session means its life cycle completes at the $j$-th session. Consequently, $U_i$ does not have $\beta_{j_1}$ for $j_1 > j$. As a result, revoked users cannot compute the subsequent session keys $SK_{j_1}$ for $j_1 > j$, as desired. This is forward secrecy.

Similarly, a user $U_i$ joined in session $j$ does not have $\beta_{j_2}$ for $j_2 < j$, although it can compute the backward keys $K^B_{m-j_2+1}$ for $j_2 < j$. This forbids $U_i$ to compute the previous session keys as desired. This is backward secrecy.

Now we will show that our construction can resist collusion required by point 3(b) of Definition 5. Let $1 \leq j_1 < j < j_2 \leq m$. For any disjoint subsets $L_1, L_2 \subset U$, where $|L_1 \cup L_2| \leq t$, no information about the session key $SK_j$, $j_1 < j < j_2$ can be obtained by the coalition $L_1 \cup L_2$, where the set $L_1$ is a coalition of users removed before session $j_1$ and the set $L_2$ is a coalition of users joined from session $j_2$. Our constructions satisfy this property as illustrated below. Secret information held by users in $L_1 \cup L_2$ and broadcasts in all the sessions do not get any information about $SK_j$ for $j_1 < j < j_2$. This is true because in the worst case, the coalition knows $S_i = \{v_1.\Phi(U_i), \ldots, v_{j_1-1}.\Phi(U_i); \beta_1, \ldots, \beta_{j_1-1}\}$ for $U_i \in L_1$, $S_i = \{v_{j_2}.\Phi(U_i), \ldots, v_m.\Phi(U_i); \beta_{s_i}, \ldots, \beta_{t_i}; \beta_{j_2}, \ldots, \beta_m\}$ for $U_i \in L_2$, and $\mathcal{B}_1, \ldots, \mathcal{B}_m$. For each session $j$, $j_1 \leq j \leq j_2 - 1$, the coalition can get backward key $K^B_{m-j+1}$ from $L_2$. However, the session key $SK_j$ is computed from the backward key $K^B_{m-j+1}$ and a random number $\beta_j$. The coalition $L_1 \cup L_2$ cannot obtain the random numbers $\beta_j$ for $j_1 \leq j < j_2$. Consequently, all the guess for $SK_j$ with $j_1 \leq j < j_2$ are equi-probable.

## 5 Complexity

- *Storage overhead:* Storage complexity of personal key for user $U_i$ with life cycle $(s_i, t_i)$ is $(t_i - s_i + t + 2)\log q$ bits.

- *Communication overhead:* The communication bandwidth for key management at the $j$-th session is $(t_j + 1)\log q$ bits, where $t_j = |W_j \cup R_j|$, $R_j \notin \Gamma$ is the set of all revoked users for sessions in and before $j$ and $W_j \subset \mathcal{U} \setminus G_j$ with minimum cardinality such that $W_j \cup R_j \in \overline{\Gamma}_0$. Here we ignore the communication overhead for the broadcast of user identities $U_i$ for $U_i \in W_j \cup R_j$, as these identities can be picked from a small finite field. In particular, if our scheme is obtained from Shamir's $(t, n)$-threshold secret sharing scheme that realises access structure defined by $\Gamma = \{A \subseteq \mathcal{U} : |A| \geq t\}$ by means of polynomial interpolation, then communication bandwidth for key management is $(t + 1)\log q$ bits.

- *Computation overhead:* The computation complexity is $2(t_j^2 + t_j)$, where $t_j = |W_j \cup R_j|$, $R_j \notin \Gamma$ is the set of all revoked users for sessions in and before $j$ and $W_j \subset \mathcal{U} \setminus G_j$ with minimum cardinality such that $W_j \cup R_j \in \overline{\Gamma}_0$. This is the number of multiplication operations needed to recover $\Phi(GM)$ by using equation (1). Considering Shamir's $(t, n)$-threshold secret sharing scheme, the computation cost for key management is $2(t^2 + t)$, which is essentially the number of multiplication operations needed to recover a $t$-degree polynomial by using Lagrange's interpolation formula.

## 6 Security analysis

Theorem 1: *Our construction is secure, self-healing session key distribution scheme with privacy, $\mathcal{R}$-revocation capability with respect to Definition 5 in our security model as described in Section 2.5 and achieve $\mathcal{R}$-wise forward and backward secrecy with respect to Definition 6 in the model.*

*Proof:* Our goal is security against coalition of users from $\mathcal{R}$. We will show that our construction is computationally secure with respect to revoked users under the difficulty of inverting one-way function, i.e. for any session $j$ it is computationally infeasible for any set of revoked users from $\mathcal{R}$ before and on session $j$ to compute with non-negligible probability the session key $SK_j$, given the *View* consisting of personal keys of revoked users, broadcast messages before, on and after session $j$ and session keys of revoked users before session $j$.

Consider a coalition of revoked users from $\mathcal{R}$, say $R_j \in \mathcal{R}$, who are revoked on or before the $j$-th session. The revoked users are not entitled to know the $j$-th session key $SK_j$. We can model this coalition of users from $\mathcal{R}$ as a polynomial-time algorithm $\mathcal{A}'$ that takes *View* as input and outputs its guess for $SK_j$. We say that $\mathcal{A}'$ is successful in breaking the construction if it has a non-negligible advantage in determining the session key $SK_j$. Then using $\mathcal{A}'$, we can construct a polynomial-time algorithm $\mathcal{A}$ for inverting one-way function $\mathcal{H}$ and have the following claim:

Claim: *Assuming a cryptographically secure PRNG, $\mathcal{A}$ inverts one-way function $\mathcal{H}$ with non-negligible probability if $\mathcal{A}'$ is successful.*

*Proof:* Given any instance $y = \mathcal{H}(x)$ of one-way function $\mathcal{H}$, $\mathcal{A}$ first generates an instance *View* for $\mathcal{A}'$ as follows: $\mathcal{A}$ randomly generates $m$ distinct number $\beta_1,\ldots,\beta_m \in GF(q)$ by using a cryptographically secure PRNG and constructs the following backward key chain by repeatedly applying $\mathcal{H}$ on $y$: $K_1^B = y, K_2^B = \mathcal{H}(y), K_3^B = \mathcal{H}^2(y),\ldots,K_j^B = \mathcal{H}^{j-1}(y),\ldots,K_m^B = \mathcal{H}^{m-1}(y)$.

$\mathcal{A}$ computes the $j$-th session key $SK_j = \beta_j + K_{m-j+1}^B$. $\mathcal{A}$ chooses at random $m$ vectors $v_1,\ldots,v_m \in GF(q)^l$. For $1 \le i \le n$, each user $U_i \in \mathcal{U}$ with life cycle, say $(s_i, t_i)$, $1 \le s_i < t_i \le m$ (which is assigned to $U_i$ by $\mathcal{A}$), receives its personal secret keys corresponding to the $k_i$ sessions $S_i = \{v_{s_i}.\Phi(U_i),\ldots,v_{t_i}.\Phi(U_i); \beta_{s_i},\ldots,\beta_{t_i}\} \in GF(q)^{2k_i}$ from $\mathcal{A}$ via the secure communication channel between them.

In this setting, $\Gamma = 2^U \setminus \mathcal{R}$ is a monotone increasing access structure of authorised users over $\mathcal{U}$. $\Gamma$ is determined by the family of *minimal qualified subsets*, $\Gamma_0$, which is called the basis of $\Gamma$. Now $R_j \in \mathcal{R}$ implies $R_j \notin \Gamma$.

Let $G_j$ be the set of all non-revoked users in session $j$. At the $j$-th session, $\mathcal{A}$ chooses a subset of users $W_j \subset \mathcal{U} \setminus G_j$ with minimal cardinality such that $W_j \cup R_j \in \overline{\Gamma}_0$. $\mathcal{A}$ then computes broadcast message $\mathcal{B}_j$ for $j = 1,\ldots,m$ as:

$\mathcal{B}_j = \{(U_k, v_j.\Phi(U_k)) : U_k \in W_j \cup R_j\} \cup \{Z_j\}$,

where $Z_j = K_{m-j+1}^B + v_j.\Phi(GM)$. Then $\mathcal{A}$ sets *View* as

$$View = \begin{cases} v_s.\Phi(U_k) \text{ for all } U_k \in R_j \text{ and } s = 1,\ldots,m; \\ \mathcal{B}_j \text{ for } j = 1,\ldots,m; \\ \beta_1,\ldots,\beta_{j-1}; \\ SK_1,\ldots,SK_{j-1} \end{cases}$$

$\mathcal{A}$ gives *View* to $\mathcal{A}'$, which in turn selects $X, \beta_{j'} \in GF(q)$ randomly, sets the $j$-th session key to be $SK_{j'} = \beta_{j'} + X$ and returns $SK_{j'}$ to $\mathcal{A}$. $\mathcal{A}$ checks whether $SK_{j'} = SK_j$. If not, $\mathcal{A}$ chooses a random $x' \in GF(q)$ and outputs $x'$.

Note that from *View*, $\mathcal{A}'$ knows $\{v_j.\Phi(U_k) : U_k \in W_j \cup R_j, 1 \le j \le m\} \cup \{v_s.\Phi(U_k) : U_k \in R_j, 1 \le s \le m\}$, $\beta_1,\ldots,\beta_{j-1}$ and at most $j-1$ session keys $SK_1,\ldots,SK_{j-1}$. Consequently $\mathcal{A}'$ has knowledge of at most $j-1$ backward keys $K_m^B,\ldots,K_{m-j+2}^B$. Observe that $SK_{j'} = SK_j$ provided

1. the guess $\beta_{j'}$ of $\mathcal{A}'$ for $\beta_j$ is correct; and
2. $\mathcal{A}'$ knows the backward key $K_{m-j+1}^B$.

The condition 1 occurs if either of the following two holds:

- $\mathcal{A}'$ is able to choose $\beta_{j'} \in GF(q)$ so that $\beta_{j'} = \beta_j$, the probability of which is $1/q$ (negligible for large $q$).

- $\mathcal{A}'$ is able to generate $\beta_{j'}$ from *View*. Note that from *View*, $\mathcal{A}'$ knows $\beta_1,\ldots,\beta_{j-1} \in GF(q)$. Observe that $\beta_1,\ldots,\beta_{j-1}$ are generated by a cryptographically secure PRNG. Thus if $\mathcal{A}'$ is able to generate $\beta_{j'}$ from the known random numbers $\beta_1,\ldots,\beta_{j-1}$, then the PRNG is insecure, leading to a contradiction.

The condition 2 occurs if either of the following two holds:

- $\mathcal{A}'$ is able to compute the $v_j.\Phi(GM)$ from *View* and consequently can recover the backward key $K_{m-j+1}^B$ as follows: $K_{m-j+1}^B = Z_j - v_j.\Phi(GM)$. From *View*, $\mathcal{A}'$ knows $\{v_j.\Phi(U_k) : U_k \in W_j \cup R_j, 1 \le j \le m\} \cup \{v_s.\Phi(U_k) : U_k \in R_j, 1 \le s \le m\}$, where $W_j \subset \mathcal{U} \setminus G_j$ has minimal cardinality with $W_j \cup R_j \in \overline{\Gamma}_0$ and will not be able to compute $v_j.\Phi(GM)$ by the property of $\Phi$. Observe that $v_j.\Phi(GM)$ is linear combination of $\{v_j.\Phi(U_k) : U_k \in B\}$ if and only if $B \in \Gamma$. Consequently, $\mathcal{A}'$ will not be able to recover $K_{m-j+1}^B$ from $\mathcal{B}_j$ as described above.

- $\mathcal{A}'$ is able to choose $X \in GF(q)$ so that the following relations hold:

$$K_m^B = \mathcal{H}^{j-1}(X)$$
$$K_{m-1}^B = \mathcal{H}^{j-2}(X)$$
$$\vdots$$
$$K_{m-j+2}^B = \mathcal{H}(X)$$

This occurs with a non-negligible probability only if $\mathcal{A}$ is able to invert the one-way function $\mathcal{H}$. In that case, $\mathcal{A}$ returns $x = \mathcal{H}^{-1}(y)$.

The above arguments show that if $\mathcal{A}'$ is successful in breaking the security of our construction, then $\mathcal{A}$ is able to invert the one-way function.

Hence our construction is computationally secure under the hardness of inverting one-way function and the security of the PRNG. This is forward secrecy. We can also prove the computational security for backward secrecy of our construction using the similar arguments as above considering a coalition of new joined users. The only difference in the proof is that this coalition of new users joined in and after session $j$ knows all the backward keys, but they do not know $\beta_1, \ldots, \beta_{j-1}$ and consequently are unable to compute the past session keys they were unauthorised to.

We will now show that our construction satisfies all the conditions required by Definition 5.

1 (a) Session key efficiently recovered by a non-revoked user $U_i$ is described in the third step of our construction.

(b) For any set $R_j \subseteq \mathcal{U}$, $R_j \in \mathcal{R}$, and any non-revoked user $U_i \notin R_j$, we show that the coalition $R_j$ knows nothing about the personal secret $S_i = (v_{s_i}.\Phi(U_i), \ldots, v_j.\Phi(U_i), \ldots, v_{t_i}.\Phi(U_i); \beta_{s_i}, \ldots, \beta_j, \ldots, \beta_{t_i})$ of $U_i$ with life cycle $(s_i, t_i)$, $1 \leq s_i \leq t_i \leq m$. For any session $j$, $U_i$ uses $v_j.\Phi(U_i)$ and $\beta_{j'}$ as its personal secret. Since the coalition $R_j \notin \Gamma$, the values $\{v_s.\Phi(U_k) : U_k \in R_j, 1 \leq s \leq m\}$ is not enough to compute $v_j.\Phi(U_i)$ by the property of $\Phi$. Moreover, the coalition $R_j$ may at most learn $\beta_1, \ldots, \beta_{j-1}$ and the probability of $R_j$ to guess $\beta_{j'}$ is negligible under the security of cryptographically secure PRNG. So it is computationally infeasible for coalition $R_j$ to learn $v_j.\Phi(U_i)$ for $U_i \notin R_j$.

(c) The session key $SK_j$ for the $j$-th session is computed from two parts: backward key $K^B_{m-j+1}$ and random number $\beta_{j'}$ where $\beta_{j'}$ are parts of personal key received from GM before or when it joins the session group and $K^B_{m-j+1} = Z_j - v_j.\Phi(GM)$ is recovered from the broadcast message $\mathcal{B}_j$. So the personal secret keys alone do not give any information about any session key. Since the initial backward seed $S^B$ is chosen randomly, the backward key $K^B_{m-j+1}$ and consequently the session key $SK_j$ is random as long as $S^B$, $K^B_1, K^B_2, \ldots, K^B_{m-j+2}$ are not get revealed. This in turn implies that the broadcast messages alone cannot leak any information about the session keys. So it is computationally infeasible to determine $Z_{i,j}$ from only personal key $S_i$ or broadcast message $\mathcal{B}_j$.

2 ($\mathcal{R}$-revocation property) Let $R_j \subseteq \mathcal{U}$, where $R_j \in \mathcal{R}$, collude in session $j$. It is impossible for coalition $R_j$ to learn the $j$-th session key $SK_j$ because the knowledge of $SK_j$ implies the knowledge of either the backward key $K^B_{m-j+1}$ or $\mathcal{B}_j$ or the knowledge of the personal secret $v_j.\Phi(U_i)$ of user $U_i \notin R_j$. The coalition $R_j$ knows the set $\{v_s.\Phi(U_k) : U_k \in R_j, 1 \leq s \leq m\}$, which is not enough to compute $v_j.\Phi(U_i)$ by the property of $\Phi$. Hence the coalition $R_j$ cannot recover $v_j.\Phi(U_i)$, which in turn makes $K^B_{m-j+1}$ appears random to all users in $R_j$. Moreover the coalition knows at most $\beta_1, \ldots, \beta_{j-1}$ and guessing $\beta_{j'}$ is negligible under the security of PRNG. Therefore, $SK_j$ is completely safe to $R_j$ from computation point of view.

3 (a) (Self-healing property) From the third step of our construction, any user $U_i$ that is a member in sessions $j_1$ and $j_2$ $(1 \leq j_1 < j_2)$, can recover the backward key $K^B_{m-j_2+1}$ and hence can obtain the sequence of backward keys $K^B_{m-j_1}, \ldots, K^B_{m-j_2+2}$ by repeatedly applying $\mathcal{H}$ on $K^B_{m-j_2+1}$. User $U_i$ also holds $\beta_{j_1+1}, \ldots, \beta_{j_2-1}$. Hence, as shown in Section 3, user $U_i$ can efficiently recover all missed session keys.

(b) Our construction can also resist $\mathcal{R}$-wise collusion. Let $1 \leq j_1 < j < j_2 \leq m$ and let $L_1, L_2 \in \mathcal{U}$ be two disjoint subsets, where $L_1$ is a set of revoked users from the group before $j_1$ and $L_2$ is a set of users who join the group from session $j_2$. Consider a coalition from $L_1 \cup L_2 \in \mathcal{R}$. We show the users in $L_1 \cup L_2$ together are not entitled to know the $j$-th session key $SK_j$ for any $j_1 \leq j < j_2 - 1$. We can model this coalition of users $L_1 \cup L_2$ as a polynomial-time algorithm $\mathcal{A}'$ that takes View as input and outputs its guess for $SK_j$. We say that $\mathcal{A}'$ is successful in breaking the construction if it has a non-negligible advantage in determining the session key $SK_j$.

$\mathcal{A}$ first generates the instance View for $\mathcal{A}'$ as follows:

$$View = \begin{cases} v_s.\Phi(U_k) \, \forall \, U_k \in L_1 \cup L_2 \text{ and } s = 1, \ldots, m; \\ \mathcal{B}_j \text{ for } j = 1, \ldots, m; \\ S^B; \\ \{\beta_1, \ldots, \beta_{j_1-1}\} \cup \{\beta_{j_2}, \ldots, \beta_m\}; \\ \{SK_1, \ldots, SK_{j_1-1}\} \cup \{SK_{j_2}, \ldots, SK_m\} \end{cases}$$

$\mathcal{A}$ gives View to $\mathcal{A}'$ which in turn selects $\beta_{j'} \in GF(q)$ randomly, sets the $j$-th session key to be $SK_{j'} = \beta_{j'} + K^B_{m-j+1}$, and returns $SK_{j'}$ to $\mathcal{A}$. $\mathcal{A}$ checks whether $SK_{j'} = SK_j$. If not, $\mathcal{A}$ chooses a random $x' \in GF(q)$ and outputs $x'$.

$\mathcal{A}'$ can compute the $j$-th backward key $K_{m-j+1}^B = H^{m-j+1}(S^B)$ because it knows $S^B$ from $View$ for $j = 1, \ldots, m$. Note that from $View$, $\mathcal{A}'$ knows random numbers $\{\beta_1, \ldots, \beta_{j_1-1}\} \cup \{\beta_{j_2}, \ldots, \beta_m\}$ and session keys $\{SK_1, \ldots, SK_{j_1}\} \cup \{SK_{j_2}, \ldots, SK_m\}$. Observe that $SK_{j'} = SK_j$ provided $\mathcal{A}'$ knows the random $\beta_{j'}$, which occurs if either of the following two conditions hold:

- $\mathcal{A}'$ is able to choose $\beta_{j'} \in GF(q)$ so that $\beta_{j'} = \beta_j$, the probability of which is $1/q$ (negligible for large $q$).
- $\mathcal{A}'$ is able to generate $\beta_{j'}$ from $View$. Note that from $View$, $\mathcal{A}'$ knows $\{\beta_1, \ldots, \beta_{j_1-1}\} \cup \{\beta_{j_2}, \ldots, \beta_m\}$ which are generated by a cryptographically secure PRNG. Thus if $\mathcal{A}'$ is able to generate $\beta_{j'}$ from the known random numbers $\{\beta_1, \ldots, \beta_{j_1-1}\} \cup \{\beta_{j_2}, \ldots, \beta_m\}$, then the PRNG is insecure, leading to a contradiction.

The above arguments show that if $\mathcal{A}'$ is successful in breaking the security of our construction, then the PRNG used to generate the random numbers by $\mathcal{A}$ is insecure. Hence our construction is computationally secure for resisting $\mathcal{R}$-coalition under the assumption that the PRNG is cryptographically secure.

We will show that our construction satisfies all the conditions required by Definition 6.

1. ($\mathcal{R}$-wise forward secrecy) Let $R_j \subseteq \mathcal{U}$, where $R \in \mathcal{R}$ and all user $U_s \in R_j$ are revoked before the current session $j$. The coalition $R_j$ cannot get any information about the current session key $SK_j$ even with the knowledge of group keys before session $j$. This is because of the fact that in order to know $SK_j$, any user $U_s \in R_j$ needs to know either $v_j.\Phi(GM)$ or $K_{m-j+1}^B$ or $\beta_{j'}$. Determining $v_j.\Phi(GM)$ requires knowledge of values $\{v_j.\Phi(U_k) : U_k \in B \text{ for some } B \in \Gamma\}$. But the coalition $R_j$ knows only the values $\{v_j.\Phi(U_s) : U_s \in R_j\}$ which is insufficient as $R_j \notin \Gamma$. Hence $R_j$ is unable to compute $SK_j$. Besides, because of the one-way property of $\mathcal{H}$, it is computationally infeasible to compute $K_{j_1}^B$ from $K_{j_2}^B$ for $j_1 < j_2$. The users in $R_j$ might know the sequence of backward keys $K_m^B, \ldots, K_{m-j+2}^B$, but cannot compute $K_{m-j+1}^B$ and consequently $SK_j$ from this sequence. Hence our construction is $\mathcal{R}$-wise forward secure. Moreover the coalition knows at most $\beta_1, \ldots, \beta_{j-1}$ and guessing $\beta_{j'}$ is negligible under the security of PRNG.

2. ($\mathcal{R}$-wise backward secrecy) Let $J_j \subseteq \mathcal{U}$, where $J_j \in \mathcal{R}$ and all user $U_s \in J_j$ join after the current session $j$. The coalition $J_j$ can not get any information about any previous session key $SK_{j_1}$ for $j_1 \leq j$ even with the knowledge of group keys after session $j$. This is because of the fact that in order to know $SK_{j_1}$, any user $U_s \in J_j$ requires the knowledge of $\beta_{j_1}$. Now when a new member $U_v$ joins the group starting from session $j+1$, the GM gives $U_v$ at most $\beta_{j+1}, \ldots, \beta_m$, together with the values $S_v = (v_{j+1}.\Phi(U_v), \ldots, v_m.\Phi(U_v)) \in GF(q)^{m-j+2}$. Hence it is computationally infeasible for the newly joined member to trace back for previous $\beta_{j_1}$ under the security of PRNG for $j_1 \leq j$. Consequently, our protocol is $\mathcal{R}$-wise backward secure. In fact, this backward secrecy is independent of $\mathcal{R}$.

## 7 Comparison

The storage overhead, communication complexity and computation cost of each user in our construction is provided in Section 5. The existing works to deal with self-healing key distribution using monotone decreasing family of revoked subset of users instead of monotone decreasing threshold structure are Dutta et al. (2008) and Saez (2004). In contrast to the family of the self-healing key distribution schemes proposed by Saez (2004), our general construction uses a different self-healing approach based on Dutta et al. (2008) which is more efficient in terms of computation and communication, yielding a family of more flexible self-healing key distribution schemes that can provide better properties. Unlike Saez (2004), the length of the broadcast message in our scheme does not depend on the history of revoked subsets of users to perform self-healing. This feature provides significant reduction in the communication cost, which is one of the main improvement of our scheme over the previous works (Staddon et al., 2002; Liu et al., 2003; Blundo et al., 2004; Hong and Kang, 2005). For simplicity, we compare a special case of our construction with the other similar schemes considering Shamir's $(t, n)$-threshold secret sharing.

If we consider a secret sharing scheme realising a specific bipartite access structure defined in the set of users, the previous self-healing mechanisms (Staddon et al., 2002; Liu et al., 2003; Blundo et al., 2004; Hong and Kang, 2005) allow to improve the efficiency of revocations of a small number of users, say less than $j$, for some positive integer $j \leq t-1$, $t$ is the threshold on the number of revoked users. This is because of the fact that in all the previous self-healing key distribution schemes, a part of the broadcast message of every session contains a history of revoked subsets of users in order to perform self-healing. This part of broadcast message has a proportional amount of information to $t-1$ in all the previous self-healing key distribution schemes, despite only two or three users must be revoked. We overcome this overhead on broadcast message length in our general construction since our self-healing mechanism does not need to send any such history.

To be more precise, let us use the bipartite access structure $\Gamma$ in Example 3 of Section 2, which cannot be realised by a vector space secret sharing (except in the threshold case $j = t - 1$), but by a linear one in which each participant is associated with two vectors instead of one. Each operation will have twice the cost of the same operation in the threshold case. In particular, the length of the personal keys is twice the length in the threshold case. This scheme is useful in case the efficiency in the revocation of small subsets has priority. Implementing this using our self-healing key distribution reduces communication overhead significantly as compared to the previous schemes.

Table 2 shows comparisons of different self-healing schemes in terms of storage, communication and computation, where $T_j = t$ if the access structure is realised by Shamir's $(t, n)$-threshold secret sharing scheme. In one hand, our construction reduces the communication complexity (bandwidth) to $O(t)$, whereas optimal communication complexity achieved by the previous schemes is $O(tj)$ at the $j$-th session. On the other hand, we achieve less computation cost. For a user $U_i$ at the $j$-th session, the computation cost is incurred by recovering all previous session keys up to the $j$-th session (worst case) by self-healing mechanism. The communication complexity and computation cost in our constructions do not increase as the number of session grows. These are the most prominent improvement of our schemes over the previous self-healing key distributions (Staddon et al., 2002; Liu et al., 2003; Blundo et al., 2004; Hong and Kang, 2005). Figures 3 and 4 respectively show the comparative summary of communication and computation costs of our scheme with the existing self-healing key distribution schemes for $j = 50$ with $m = 100$ and $q = 67$.

**Table 2** Comparison among different self-healing key distribution schemes in $j$-th session ($k_i = t_i - s_i + 1$, where $(s_i, t_i)$ is the life cycle assigned to user $U_i$ by the GM; $T_j$ is a threshold on the number of revoked users which depend on the monotone decreasing access structure; and $t$ is the maximum number of revoked users)

| Schemes | Storage Overhead | Communication Overhead | Computation Overhead |
| --- | --- | --- | --- |
| Staddon et al., 2002 | $(m - j + 1)^2 \log q$ | $(mt^2 + 2mt + m + t) \log q$ | $2mt^2 + 3mt - t$ |
| Liu et al., 2003 | $2(m - j + 1) \log q$ | $[(m + j + 1)t + (m + 1)] \log q$ | $mt + t + 2tj + j$ |
| Blundo et al., 2004 | $(m - j + 1) \log q$ | $(2tj + j) \log q$ | $2j(t^2 + t)$ |
| Hong and Kang, 2005 | $(m - j + 1) \log q$ | $(tj + j - t - 1) \log q$ | $2tj + j$ |
| Dutta et al., 2008 | $(m - j + 2) \log q$ | $(T_j + 1) \log q$ | $2(T_j^2 + T_j)$ |
| Our Construction | $(k_i + 1) \log q$ | $(T_j + 1) \log q$ | $2(T_j^2 + T_j)$ |

**Figure 3** Comparison of communication bandwidth with $m = 100$, $q = 67$ and $j = 50$ (see online version for colours)

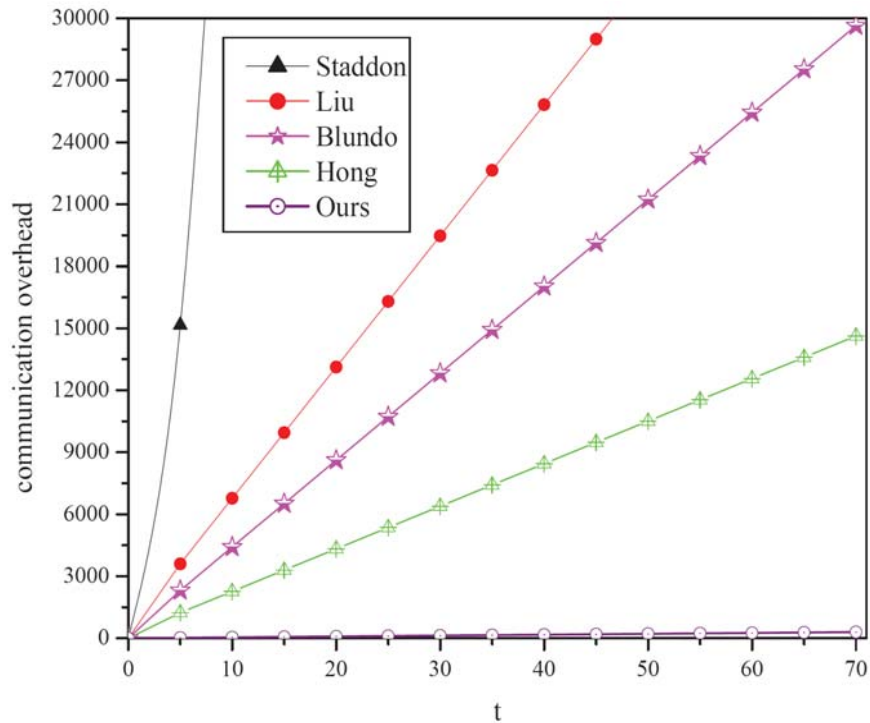

**Figure 4** Comparison of computation overhead with $m = 100$, $q = 67$ and $j = 50$ (see online version for colours)

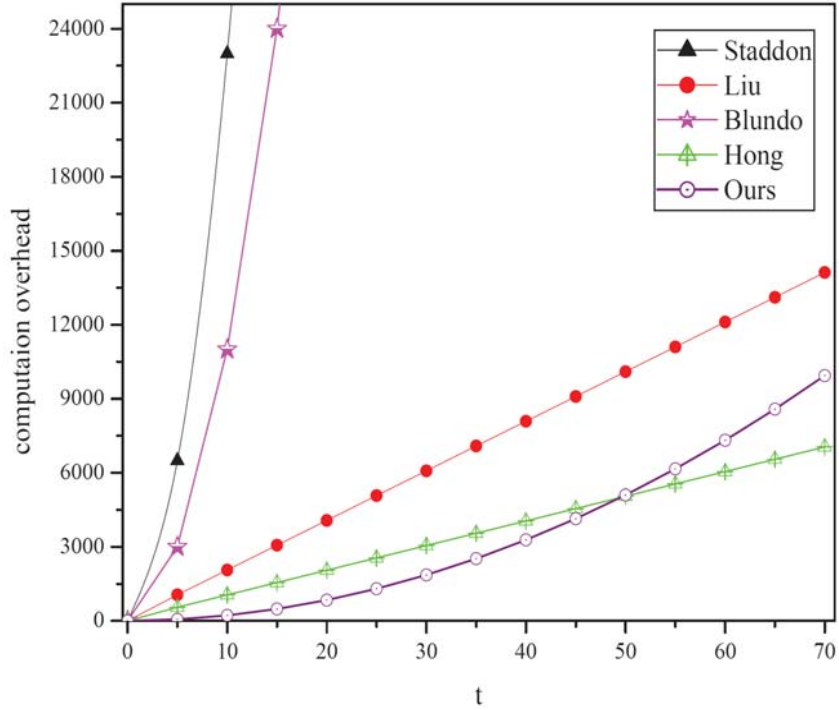

As mentioned earlier, our construction is also based on Dutta et al. (2008). However, we have the following subtle differences:

(a) No forward key chain is used in our construction unlike Dutta et al. (2008).

(b) In contrary to Dutta et al. (2008), each user $U_i$ in our construction is pre-assigned a life cycle $(s_i, t_i)$ by the GM. This means user $U_i$ can participate in $k_i = t_i - s_i + 1$ sessions and cannot revoke before session $t_i$ is over. Consequently, the GM no longer need to use computation-intensive traitor tracing techniques to keep track of compromised users to perform revocation.

(c) In contrast to Dutta et al. (2008), we have been able to resist collusion attack in our constructions by using pre-selected random numbers $\beta_1, \ldots, \beta_m$ (fixed) as part of users' secret keys. A user $U_i$ with life cycle $(s_i, t_i)$ is given only $k_i = t_i - s_i + 1$ values $v_{s_i} \cdot \Phi(U_i), \ldots, v_{t_i} \cdot \Phi(U_i)$ and the additional values $\beta_{s_i}, \ldots, \beta_{t_i}$ as part of its secret key by the GM via a secure communication channel between them at the initial set-up. As compared to Dutta et al. (2008), we get increased storage for our scheme if $k_i > \frac{m - j + 1}{2}$. The communication and computation costs for our scheme are the same as in Dutta et al. (2008).

(d) Unlike previous self-healing key distribution schemes, revoked users may join at later sessions with new identities without violating any security.

Let us now note down a few points in the following remarks.

*Remark 1:* We assume that the number of sessions ($m$) is fixed. The system fails when all $m$ sessions are exhausted or the set of revoked users for sessions in and before the current session becomes an element of $\Gamma$ for our general construction. Once the parameters associated with the sessions are run out, a new set-up process needs to be carried out. The number of sessions $m$ may be chosen large enough to prevent re-initialisation if the users have no storage constraints. But the communication bandwidth and computation for key management grows linearly with the size of the set of revoked users $R$. So $|R|$ cannot be too large in order to eliminate re-initialisation. However, this is a common problem with all the existing self-healing key distribution schemes with $R$-revocation.

*Remark 2:* The broadcast message $\mathcal{B}_j$ in the $j$-th session contains IDs (or indices of IDs) of (revoked) users. There is no privacy issue while broadcasting these valid user IDs to the network. We ignore the communication overhead for the broadcast of the set of IDs, because user IDs can be selected from a small finite field (Hong and Kang, 2005).

*Remark 3.* We emphasise that during the set-up phase, the personal secret keys of each user corresponding to $m$ sessions and corresponding $\beta_j$'s (generated by running a PRNG) associated with user's pre-assigned life cycle are delivered by the $GM$ to the corresponding user through a secure communication channel between them. The same is done while a new user joins the group. For instance, in case of a pay TV channel, a subscriber may obtain a set top box from the cable operator (GM) during its registration (set-up) phase, which stores the secret values corresponding to this subscriber.

*Remark 4.* As in the existing self-healing key distribution schemes, we also consider a typical wireless ad hoc network with a fixed group manager, instead of a general one without group managers and the revoked users are never allowed to rejoin the group. There are several special applications with this type of set-up. We handle key management in this set-up and incorporate self-healing and revocation (up to a threshold value *t*) properties with reduced overheads. This is by no means a trivial task. The idea of using secret sharing or revocation polynomial in designing key distribution schemes is not new in the literature. Our contribution is in introducing an interesting anti-collusive self-healing mechanism by applying one way functions which is effective in saving overheads and assigning pre-selected life cycle to each user during its set-up. One point not discussed above is how to handle the situation where all session parameters have been consumed and a node has no idea about the situation and wants to return to an old session. We may control this using session identities which are session specific and may be implemented by keeping a counter by each node. The counter is incremented by the node when a new session starts and the node is not revoked yet.

*Remark 5.* The group manager decides when a user should get revoked from the system. A user may get malicious any time during a legitimate session. Unless the group manager detects this fact, the user cannot be revoked by the group manager from the system. Thus the group manager has to keep track of compromised users using some special treatment such as traitor tracing, which might be expensive. The risk still remains in the system to have a malicious user until the group manager detects certain misbehaviour of the user. This is a common problem with all the existing self-healing key distribution schemes. In our pre-arranged life-cycle based approach, there is no need for the group manager to use expensive traitor tracing algorithms in handling compromised nodes. The selection of a user's life cycle is pre-determined by the group manager. The group manager believes that a user behaves honestly and will not get compromised during its life cycle. In our setting, the group manager pre-selects the session of revocation for a user during user's set-up phase by assigning the user a pre-arranged life cycle. The user is revoked from the system by the group manager once its life cycle is over irrespective of user gets compromised or not. The joining session can be selected by the user. Our designs allow a revoked user to join at a later session with new identity and a new life cycle starting from its new joining session.

*Remark 6.* Our security model addresses the inside attacks only and the security analysis of our scheme is in this security framework. However, additional security measures should be provided for the group manager to block the intruders from compromising it, thereby preventing the group manager itself to fail under attack. One of the ways to counter such an event is frequent changing of the group manager. Many practical applications of sensor networks require their cluster head roles to be undertaken by different nodes at different time instants (though the aim is mainly to conserve energy). Thus the activities of each such manager node can be voted upon by the remaining members, and the re-assignment of the node to the manager position is contingent on receiving at least a threshold number of votes. Thus, the damage caused to the network can be minimised over long evaluation times, and in the asymptotic case, the malicious group manager will soon be disallowed from resuming the governing activity over the network. It is an open challenge to identify the interval in which member votes can be counted, given the healing time of a proposed method. We shall study this further in future work.

# 8 Conclusion

We introduce the collusion resistance property to the generalised self-healing key distribution proposed by Dutta et al. (2008) using a pre-arranged life-cycle based approach. Our set-up allows each user to choose its joining session at its will, but the session for its revocation is pre-selected by the group manager. In contrast to polynomial based schemes, our scheme has realised a general monotone decreasing access structure for the family of subsets that can be revoked instead of a threshold one. This provides more flexible performance for self-healing key distribution and would suit various wireless network environments. Several innovative business models allow contractual subscription or rental by the service provider for the scalability of business and do not allow the user to revoke before its contract is terminated. Our pre-determined life cycle based key distribution scheme is suitable for such applications. The proposed scheme provides better efficiency in communication and storage as compared to the existing approaches. Most important of all, our scheme can resist collusion between the newly joined users and the revoked users besides forward and backward secrecy. The scheme has been properly analysed in an appropriate security model and is proven to be computationally secure. Moreover, rejoining of revoked users can be done in our scheme at later sessions with new identities without compromising security, unlike the existing self-healing schemes.